# Integration of Efficacy Biomarkers Together with Toxicity Endpoints in Immune-Oncology Dose Finding Studies

Yiding Zhang[1], Zhixing Xu[2], Hui Quan[2], Ji Lin[1]

[1]Biostatistics and Programming, Sanofi, 450 Water Street, Cambridge, MA 02141

[2]Biostatistics and Programming, Sanofi, 55 Corporate Drive, Bridgewater, NJ 08807

**Abstract**

The primary objective of phase I oncology studies is to establish the safety profile of a new treatment and determine the maximum tolerated dose (MTD). This is motivated by the development of cytotoxic agents based on the underlying assumption that the higher the dose, the greater the likelihood of efficacy and toxicity. However, evidence from the recent development of cancer immunotherapies that aim to stimulate patients' immune systems to fight cancer challenges this assumption, particularly further escalation after certain dose level might not necessarily increase the efficacy. Dose escalation study of molecular targeted agents (MTA) often does not only rely on the safety profile. In this paper, we propose a simple and flexible model that uses multivariate Gaussian latent variables to integrate toxicity endpoint and efficacy biomarker. This model can be easily extended to incorporate additional immune biomarkers. By simultaneously considering multiple outcomes, the proposed method is better at identifying the biologically optimal dose, which results in better decision-making. Simulation studies showed that the proposed method has desirable operating characteristics by determining the target dose with an optimal risk-benefit trade-off. We have also implemented our proposed method in a user-friendly R Shiny tool.

# 1. Introduction

In the traditional phase I oncology studies, the goal is generally to find the maximum tolerated dose (MTD). The safety is usually measured by the rate of dose limiting toxicity (DLT), which is "traditionally defined by grade 3/4 non-hematological or grade 4 hematological toxicity at least possibly related to the treatment, occurring during the first cycle of treatment" with some adjustments (Paoletti et al. 2014). Methods for oncology dose-escalation trials that focus on toxicity alone fall into two broad classes: rule-based designs including the classic 3 + 3 design and its variations, and model-based designs including continual reassessment method (CRM) (O'Quigley et al. 1990) dose escalation with overdose control (EWOC) (Babb et al. 1998), and Bayesian logistic regression model (BLRM) (Neuenschwander et al. 2008). In those standard phase I designs, it is assumed that probability of toxicity and efficacy increase as dose level increases. This assumption could be true for cytotoxic agents. However, with the changing landscape of oncology drug development. Molecular targeted agents (MTA) have gained popularity and may present different dose-efficacy and dose-toxicity curves from those for cytotoxic agents. Recently, the US Food and Drug Administration (FDA) encourage that "…sponsors perform dose-finding studies to evaluate exposure-response, efficacy, and safety and inform dose selection for registrational trials." (Blumenthal et al. 2022). Known as Project Optimus, more attentions are put in the early phase dose-finding studies to ensure the dose selected can provide optimal treatment for patients with cancer (Blumenthal et al. 2022; Shah et al. 2021). In such case, the traditional designs whose goals are to identify the MTD will not be suitable, and the objective of the immunotherapy trial will be to identify the biologically optimal dose (BOD), where BOD is defined as the dose yielding the highest risk-benefit trade off. Various designs have been proposed to incorporate both toxicities and efficacies in the dose finding studies, such as EffTox (Thall and Cook 2004), TriCRM (Zhang et al. 2006), EBE-CRO (Colin et al. 2017), etc. Second, the immunotherapy is an innovative treatment method that stimulates a patient's immune system to fight cancer and can generate a number of immune responses, and immune/efficacy biomarkers can provide information about the biological efficacy of the immunotherapy in activating the immune system. The immune response is often associated with the treatment effect of the immunotherapy (Galon et al. 2006; Guo, Li, and Yuan 2018). Third, it is fairly common that toxicity and surrogate efficacy/immune biomarker data may be obtained quickly as compared with primary efficacy data. One possible solution is to simply replace the primary efficacy data with single or multiple biomarkers' data into the dose finding analysis. Thus, more sophisticated but easy to be implemented methods are needed to properly utilize all available information.

We developed a novel phase I/II design to identify the biologically optimal dose for immunotherapy. In the design, we can easily integrate toxicity endpoints with efficacy/immune biomarkers. For toxicity, we consider variety of dose-toxicity curves including both early and late onset toxicity; For efficacy/immune biomarkers, we consider binary or ordinal efficacy endpoint, and the extension to continuous efficacy/immune biomarker endpoint is straight forward. We have also implemented our method in R Shiny for easy application.

A few designs have been proposed recently to incorporate the immune response together with efficacy and toxicity data, such as the integrated phase I/II design proposed by Liu et al. (2018), SPIRIT (Guo et al. 2018), BDFIT (Guo, et al, 2018), etc. The proposed method can be more flexible. While in both BDFIT and SPIRIT, the efficacy endpoint considered is progression free survival (PFS), these two designs have different utility and are suitable for different trial settings based on the study needs.

The paper is organized as follows. In Section 2, we introduce the notation and explain the proposed methodology. Then we conduct simulations in Section 3 to examine the operating characteristics and provide discussions in Section 4.

## 2. Method

**2.1 Probability Models**

Assume that there are $J$ doses under study. Let $\text{dose}_k$ be the dose given to patient k and let $d_k = \log(\text{dose}_k/\text{reference dose})$, where the reference dose can be the maximum planned dose and standardization of $d_k$ can also be implemented to improve model performance. Both logarithm transformation and original format of relative dose can be used, but we find that logarithm transformation has better performance especially when dose level difference is large. Let $Y_{Tox}$ denote the binary toxicity endpoint. If we observed toxicity for the patient, $Y_{Tox}$ equals 1. Otherwise, $Y_{Tox}$ equals 0. Let $Y_{Eff}$ denote the ordinal efficacy endpoint based on the best overall response. Let $Y_{Eff}$ equals 2, 1, and 0 if patient has relative high-level, medium, and low or no expression/measurement of efficacy biomarker. For example, Cytokeratin 18 (CK) cleaved (M30) and intact (M65) ELISAs can be used to describe apoptosis epithelial and endothelial cells (Greystoke et al. 2008).

Let $Y_{Imm}$ denote the biomarker response. The immune response is often scored as a binary variable. Let $Y_{Imm}$ equals 1 if there is a biological meaningful response (for example, CD8+ T-cells increases by a certain percentage), and 0 otherwise.

Let $p^{Tox} = \Pr(Y^{Tox} = 1|d)$, i.e., the probability of having a DLT at a given dose. We consider Probit regression with latent variable $Z^{Tox} = \alpha_1 + \beta_1 \cdot d + \varepsilon$,

$$\Pr(Y^{Tox} = 1|d) = \Phi\{E(Z^{Tox}|d)\},$$

where $\varepsilon \sim N(0,1)$, $E(\cdot)$ is expectation operator, and $\Phi(\cdot)$ is CDF of standard normal distribution. Then $Y^{Tox}$ can be viewed as an indicator of whether this latent variable is positive:

$$Y^{Tox} = \begin{cases} 0, & Z^{Tox} \leq 0 \\ 1, & otherwise \end{cases}.$$

This leads to $\Pr(Y^{Tox} = 1|d) = \Pr(\alpha_1 + \beta_1 \cdot d + \varepsilon > 0) = \Phi(\alpha_1 + \beta_1 \cdot d)$.

The parameters $\alpha_1$ and $\beta_1$ are typically estimated by maximum likelihood, and the corresponding likelihood function is

$$L(\alpha_1, \beta_1|D_n) = \prod_{i=1}^{n} [\Phi(\alpha_1 + \beta_1 \cdot d_i)]^{Y_i^{Tox}} [1 - \Phi(\alpha_1 + \beta_1 \cdot d_i)]^{1-Y_i^{Tox}}$$

where $D_n = \{(Y_i^{Tox}, d_i), i = 1, \dots, n\}$ be the observed data with $Y_i^{Tox}$ be the DLT status for the $i$-th patient and $d_i$ be the log transformed dose given to the $i$th patient.

Similarly, for the ordinal efficacy biomarker endpoint, we can take a latent variable approach. That is, let $Z^{Eff}$ denote the continuous latent variable that related to $Y^{Eff}$ for a certain dose level $d$ as follows

$$Z^{Eff} = \alpha_2 + \beta_2 \cdot d + \gamma_2 \cdot d^2 + \epsilon$$

$$Y^{Eff} = \begin{cases} 0, & Z^{Eff} < 0 \\ 1, & 0 \leq Z^{Eff} < \zeta \\ 2, & Z^{Eff} \geq \zeta \end{cases}$$

where $\zeta$ is the unknown cut point for efficacy and $\epsilon \sim N(0,1)$. $Z^{Eff}$ can be interpreted as the patient's latent trait. When $Z^{Eff}$ passes certain threshold, certain clinical outcomes, such as CR or PR, are observed. We assume that $[Z^{Tox}, Z^{Eff}|d]$ follows a bivariate normal distribution for a certain dose level $d$

$$Z = [Z^{Tox}, Z^{Eff}]'$$

$$Z \sim N\left(\begin{bmatrix} \eta^{Tox} \\ \eta^{Eff} \end{bmatrix}, \Sigma\right),$$

where $\eta^{Tox} = E(Z^{Tox}|d)$, $\eta^{Eff} = E(Z^{Eff}|d)$, and $\Sigma$ is a 2×2 covariance matrix with variance components of 1 and correlation $\rho$. One can easily extend above bivariate normal distribution to multivariate normal distribution to incorporate more biomarkers information into the model, e.g., study with additional biomarker, one can define the latent variable as follow,

$$Z \sim N\left(\begin{bmatrix} \eta^{Tox} \\ \eta^{Eff} \\ \eta^{Bio} \end{bmatrix}, \Sigma\right),$$

where $\eta^{Bio} = E(Z^{Bio}|d)$, $\Sigma$ is a 3×3 covariance matrix with variance components of 1 and correlation $\rho$.

The quadratic terms in the models for efficacy/ immune biomarker are used to accommodate the possibility that the efficacy and/or immune response may not monotonically increase with dose. $\gamma_2 < 0$ is required to make the dose-efficacy curve to be realistic. When $\gamma_2 < 0$ and $\beta_2 > 0$, the dose-efficacy curve becomes monotonically increasing. The proposed model is easy to implement without need to specify trade off contour as in EffTox (Thall and Cook 2004) and additional parameters for copula joint distribution as in TriCRM (Zhong, Koopmeiners, and Carlin 2012).

For simplicity, the latent variable is assumed to follow a bivariate normal distribution. To specify the priors for $\theta = (\alpha_1, \alpha_2, \beta_1, \beta_2, \gamma_2, \rho, \zeta)$ in the regression models, the general idea is that a typical change in a covariate is unlikely to lead to a dramatic change in the probability of the response variable. Given the priors and the interim data, we can calculate the posterior distribution. Let $N$ denote the trial sample size, and let $Y_i = (Y_i^{Tox}, Y_i^{Eff})$ denote the observed outcome for the i-th patient and $d_i$ the assigned dose, where $i = 1, ..., N$. Assume that we have k patients at a certain interim dose escalation time, and let $D_k = (Y_1, Y_2, ..., Y_k)$ be the observed data for the k patients. The likelihood for the k patients is $L(D_n|\theta) = \prod_{i=1}^{n} L(Y_i|d_i, \theta)$. Let $p(\theta)$ denote the joint prior distribution of $\theta$. The joint posterior distribution based on the k patients is $p(\theta|D_n) \propto L(D_n|\theta)p(\theta)$.

## 2.2 Control the Risk of Potential Overdosing

(Babb et al. 1998) proposed EWOC method to accelerate the dose escalation by introducing Bayesian feasible level $1 - \delta$ such that the predicted proportion of patient who is treated at a dose level that is "over toxic" equals to $\delta$. The over toxic dose levels are defined as dose levels that exceed the MTD which is defined by the maximum dose level that does not exceed certain amount of probability of DLT. This predicted proportion of "over-dosed" patients for a certain dose level can be estimated by the posterior probability of DLT that exceed the predefined MTD at that dose level. For probability model without latent variable, the posterior probability of DLT can be mapped directly from the posterior samples of the tolerance distribution. Examples of common tolerance distributions are

$$F(Z) = \begin{cases} \Phi(Z), & \text{Probit model} \\ \frac{e^Z}{1+e^Z}, & \text{logistic model} \end{cases},$$

where $Z$ is generalization of $Z^{Tox}$. After acquiring the posterior samples for parameter $\alpha$ $and$ $\beta$, one can obtain a posterior sample for $Z^* = \{z_1^*, z_2^*, ...\}$ through the linear combination with dose defined in section 2.1, e.g., $z_1^* = \alpha_1^* + \beta_1^* \cdot d + \varepsilon$ Plugging the posterior sample $Z^*$, a posterior sample of the probability of DLT can be given by

$F(Z^*)$. The posterior mean of the probability of DLT for Probit model with or without latent variable is the same, because $E(Z^*_{latent}) = E(Z^*_{no\ latent} + \varepsilon) = E(Z^*_{no\ latent})$. However, the variance of the posterior samples of probability of DLT from model with latent variable would carry extract randomness from the error term $\varepsilon$, thus the variance of posterior sample of the probability of DLT would be greater than the variance from Probit model without latent variable. Hereafter, for a fixed overdose control level, larger the variance of posterior sample, greater the predicted proportion of patient would be treated at the "over toxic" dose level. Therefore, the dose level with posterior probability of DLT which has large variance has less chance to be selected as MTD.

To show that the control of risk of overdosing in Probit model is different, we draft a concise proof as follow. For Probit model without latent variable, assume that $\alpha^*$ and $\beta^*$ are random variables from their posterior distribution, then $Var\{P(Y = 1|\alpha^*, \beta^*, d)\} = Var\{\Phi(\alpha^* + \beta^* d)\} = Var\left[\frac{1}{2}\left\{1 + \text{erf}\left(\frac{\alpha^* + \beta^* d}{\sqrt{2}}\right)\right\}\right] = \frac{1}{4} Var\{\text{erf}\left(\frac{\alpha^* + \beta^* d}{\sqrt{2}}\right)\}$, where erf $(\cdot)$ is the error function for standard normal distribution defined as erf $(x) = 2\Phi(\sqrt{2}x) - 1$. For Probit model with latent variable, the variance of the probability of DLT can be given by $Var\{P(Y = 1|\alpha^*, \beta^*, d)\} = Var\{\Phi(\alpha^* + \beta^* d + \varepsilon)\} = \frac{1}{4} Var\{\text{erf}\left(\frac{\alpha^* + \beta^* d + \varepsilon}{\sqrt{2}}\right)\}$. It is obvious that regardless of the variance of posterior samples of $\alpha$ and $\beta$, there is always a fixed source of variance that comes from the randomness of $\varepsilon$. Therefore, traditional way of controlling the overdose would not provide similar tolerance level for the risk of overdosing in Probit model with latent variable but would be too conservative for dose escalation.

Since there is no closed form for the error function, we cannot derive an exact formula for the variance to investigate how much more probability should we consider to achieve at least similar amount tolerance level for the risk of over dosing as in the model without latent variable. One could utilize simulation to estimate the posterior probability of DLT that exceed a certain MTD assuming that we know the estimated mean posterior probability of DLT. For example, generate a large number of random samples from $\Phi\{\Phi^{-1}(Z_\pi) + \varepsilon\}$, where $\varepsilon \sim N(0,1)$, $\phi(Z_\pi) = \pi$, $\pi \in (0,1)$. One can use the mean of random sample as the estimated mean posterior probability of DLT and proportion of random sample that exceed certain pre-defined MTD e.g., 0.3 as the potential risk of overdosing. The risk overdosing control level can be chosen through simulation according to investigator's need or study design. For example, the classic overdose control level of 0.25 in EWOC or BLRM could be too conservative in Probit model with latent variable if the target MTD is 0.3. Simulation shows that 0.4 is a good candidate in Probit model with latent variable if target MTD is around 0.3.

**2.3 Dose Finding Rules**

The dose escalation is conducted as following steps:

1) Define target DLT rate interval as $[T_l, T_u]$ or simply an upper bound $T_u$ and a threshold for efficacy/immune biomarker. The choice of controlling target DLT rate can vary by study specific need. In most cases, dose escalation in early oncology studies is still aiming to find MTD, so that we can choose to use target DLT rate interval rather than a single upper bound to make the dose finding safety driven. If the study is trying to find the biologically optimal dose with high efficacy while controlling the DLT rate to be below certain level and as low as possible, a single upper bound of DLT rate can be used. Because a good and reliable efficacy biomarker is not often being well studied especially for first in human trials, a clinically meaningful lower bound for efficacy biomarker should be determined carefully, especially when the single upper bound for DLT rate being chosen.

2) Assign a cohort of size $c$ (for example, 3) to a single dose level at a time, starting from the lowest dose level. Dose levels cannot be skipped during dose escalation.

3) Given the accumulated data at every interim point in the trial, update posterior distributions using the Bayesian MCMC algorithm. Find $j^*$ whose posterior probability of DLT rate in $[T_u, 1)$ is less than a cutoff, say 0.4. In practice, we will update the posterior distributions for dose escalation decision when at least 3 patients in the current cohort have been followed up for the minimum observation period (for example, 1 month).

4) Calculate the dose level with the largest posterior probability in target area based on toxicity within the target interval or less than a cutoff and efficacy /immune biomarker above a threshold, denote as $j_{target}$.

5) If $j_{recommend} > j_{curr}$, then assign $j_{curr} + 1$ dose level to the next cohort. If $j_{recommend} = j_{curr}$, then assign $j_{recommend}$ dose level to the next cohort. If $j_{recommend} < j_{curr}$, then assign $j_{curr} - 1$ dose level to the next cohort.

6) Repeat Step 2) to 4) until the stopping rule is met. When trial is stopped, the $j_{recommend}$ is the target dose level for next stage of study e.g., Phase II expansion study. If the lowest dose is too toxic, trial will stop, and no target dose will be recommended for next stage of study. The stopping rules can be based on the study needs. For example, we can stop the trial if a pre-defined sample size has reached, or we can stop the escalation if a dose has been tested twice and is recommended again.

Due to the choice of threshold, the target dose level from dose escalation study would be those dose levels that secure the safety profile and with efficacy biomarker response as high as possible. One could also consider the biomarkers that indicating the safety risk. In such cases, an upper bound can be used for safety biomarker to secure the safety profile.

## 3. Simulation Studies

We assess the operating characteristics in this section via simulation studies. Following the choices of doses in a motivating trial for developing the proposed design, we set nine doses (mg) as follows: 60, 75, 90, 105, 120, 135, 150, 165, 180. The maximum sample size is 54 with a cohort size of 3. We first consider two outcomes $\{Y^{Tox}, Y^{Eff}\}$: $Y^{Tox}$ to be binary with $Y^{Tox} = 1$ for DLT and $Y^{Tox} = 0$ for non-DLT and $Y^{Eff}$ to be ordinal with values of $\{0,1,2\}$ for {non-response, mild response, high response} respectively. The target toxicity interval $[T_l, T_u] = [0.16, 0.33]$, efficacy/immune biomarker response ($Y^{Eff} = 1\ or\ 2$) lower bound $\theta_E = 0.2$. The potential risk of overdosing is controlled at 0.4. We assume that by the time of making the escalation decision, we can have the results for efficacy/immune biomarker measurement ready. We generate 1000 data sets for each scenario in this simulation study. The priors for $\theta = (\alpha_1, \alpha_2, \beta_1, \beta_2, \gamma_2, \rho, \zeta)$ can be specified as follows: $\alpha_1 \sim N(-1, 1.25)$, $\alpha_2 \sim N(-0.5, 1.25)$ $\beta_1 \sim N(0.5, 1.25)$, $\beta_2 \sim N(0.7, 1.25)$ $\gamma_2 \sim N(-0.4, 1)$, $\zeta \sim U(0, 6)$ to represent a generally reasonable guess for the DLT rate and efficacy/immune biomarker response rate with wide confidence interval in

Figure 1 and Figure 2. The covariance matrix $\Sigma$ has a Wishart distribution with an identical scale matrix and degree freedom of 3.

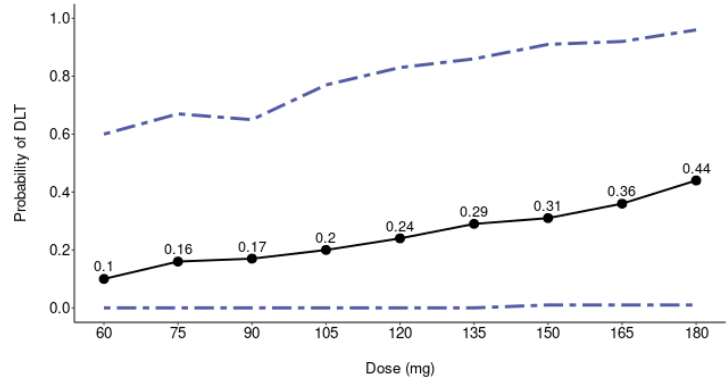

**Figure 1:** Prior DLT rate

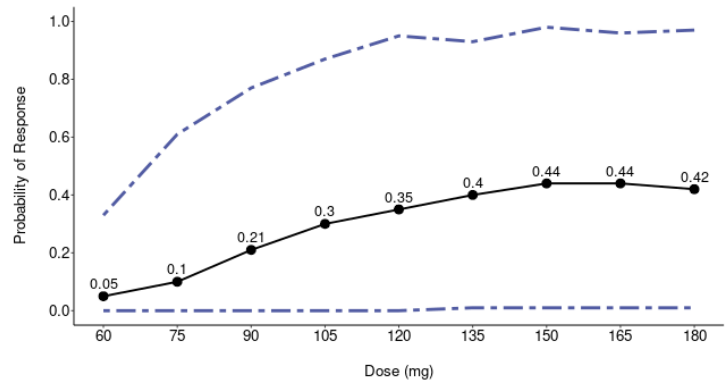

**Figure 2:** Prior Efficacy/Immune biomarker response rate

The toxicity and efficacy/immune biomarker responses are simulated from Bernoulli distribution with marginal probability of DLT and probability of efficacy/immune biomarker is responding in Table 1. The probabilities in bold font are the ones that within the target toxicity interval or above the lower threshold for efficacy/immune biomarker. Notice that toxicity settings are the same for scenario 1 - 4 to evaluate the marginal gain from classic toxicity-based dose-finding design, using the same target DLT rate interval and over toxicity control level, we compare the simulation performance with BLRM method and proposed latent Probit regression model with toxicity only. For dose-toxicity and dose-efficacy/immune biomarker response relations, we consider one and two dose levels that are within the target toxicity interval and dose- efficacy/immune biomarker response to be monotonically increasing, increasing then plateau, and bell shaped. Key features of these 6 scenarios are summarized in Table 2.

**Table 1:** Simulation Scenarios on probability of DLT and probability of efficacy/immune biomarker

| Dose | 60 | 75 | 90 | 105 | 120 | 135 | 150 | 165 | 180 | Target Dose Level |
|---|---|---|---|---|---|---|---|---|---|---|
| Dose Level | 1 | 2 | 3 | 4 | 5 | 6 | 7 | 8 | 9 | |
| Scenario 1 | | | | | | | | | | |
| Prob DLT | 0.01 | 0.05 | 0.10 | **0.18** | **0.27** | 0.38 | 0.5 | 0.55 | 0.7 | 4, 5 |
| Prob Eff | 0.05 | 0.1 | 0.18 | **0.22** | **0.23** | **0.24** | **0.25** | **0.26** | **0.27** | |
| Scenario 2 | | | | | | | | | | |
| Prob DLT | 0.01 | 0.05 | 0.10 | **0.18** | **0.27** | 0.38 | 0.5 | 0.55 | 0.7 | 4, 5 |
| Prob Eff | 0.05 | 0.1 | 0.18 | **0.38** | **0.4** | **0.42** | **0.44** | **0.45** | **0.46** | |
| Scenario 3 | | | | | | | | | | |
| Prob DLT | 0.01 | 0.05 | 0.10 | **0.18** | **0.27** | 0.38 | 0.5 | 0.55 | 0.7 | 4, 5 |
| Prob Eff | 0.05 | 0.1 | 0.18 | **0.28** | **0.4** | **0.42** | **0.44** | **0.45** | **0.46** | |
| Scenario 4 | | | | | | | | | | |
| Prob DLT | 0.01 | 0.05 | 0.10 | **0.18** | **0.27** | 0.38 | 0.5 | 0.55 | 0.7 | 4, 5 |
| Prob Eff | 0.05 | 0.15 | **0.25** | 0.38 | **0.25** | **0.2** | **0.19** | **0.19** | **0.19** | |
| Scenario 5 | | | | | | | | | | |
| Prob DLT | 0.01 | 0.05 | 0.08 | 0.12 | **0.18** | **0.27** | 0.38 | 0.5 | 0.55 | 5, 6 |
| Prob Eff | 0.05 | 0.15 | **0.25** | 0.38 | **0.25** | **0.2** | 0.19 | 0.19 | 0.19 | |
| Scenario 6 | | | | | | | | | | |
| Prob DLT | 0.01 | 0.05 | 0.10 | 0.14 | **0.25** | 0.35 | 0.5 | 0.55 | 0.7 | 5 |
| Prob Eff | 0.05 | 0.1 | 0.18 | **0.25** | **0.38** | **0.28** | **0.2** | 0.19 | 0.19 | |

**Table 2:** Simulation scenarios' key features

| | **Target Toxicity** | **Efficacy Biomarker Curve** | **Key Features** |
|---|---|---|---|
| Scenario 1 | Two candidates | Plateau, Similar for target doses | Response rate just above cut off |
| Scenario 2 | Two candidates | Plateau, Similar for target doses | High response rate |
| Scenario 3 | Two candidates | Plateau, Monotone for target doses | Increasing response rate for target doses |
| Scenario 4 | Two candidates | Bell shape, Peak in mid dose level | Decreasing response rate for target doses |
| Scenario 5 | Two candidates | Bell shape, Peak in early dose level | Decreasing response rate for target doses |
| Scenario 6 | One candidate | Bell shape | Target dose with highest response |

For two outcomes, we summarize the simulation results in terms of percentage of dose level selected as target dose (mean enrolled patient number) in Table 3. BLRM, Probit model with latent variable for toxicity only and joint modeling toxicity and efficacy/immune biomarker utilize same stopping rule described in 2.4. We compare our design that considers toxicity and primary efficacy biomarker with a popular EffTox design that utilizes a trade-off contour for dose recommendation. The EffTox(v5.2.1) software is available from MD Anderson Cancer Center biostatistics software website. Simulation results for EffTox were evaluated using fixed number of 54 patients. Parameters for EffTox are described in supplementary materials.

**Table 3:** Simulation results for two outcomes

| Dose Level | 1 | 2 | 3 | 4 | 5 | 6 | 7 | 8 | 9 | None[1] | Target Doses | Over-toxic doses |
|---|---|---|---|---|---|---|---|---|---|---|---|---|
| *Scenario 1* | | | | | | | | | | | | |
| BLRM | 0(3) | 2(3.2) | 8(3.5) | 30(4.2) | 39(4.2) | 16(2.6) | 4(1.3) | 1(0.4) | 0(0.1) | 0 | 69(8.4) | 21(4.4) |
| Probit-tox only | 0(3) | 0(3) | 2(3.1) | 24(4) | 46(4.6) | 24(3.1) | 4(1) | 0(0.2) | 0(0) | 0 | 70(8.6) | 28(4.3) |
| Probit-joint | 0(3) | 1(3) | 5(3.3) | 28(4.1) | 46(4.5) | 17(2.5) | 2(0.8) | 0(0.1) | 0(0) | 1 | 74(8.6) | 19(3.4) |
| EffTox | 3(5.1) | 15(7.5) | 41(13.4) | 27(11.8) | 8(6.9) | 2(4.1) | 2(2.6) | 1(1.3) | 1(0.9) | 1 | 35(18.7) | 6(8.9) |
| *Scenario 2* | | | | | | | | | | | | |
| Probit-joint | 0(3) | 0(3) | 5(3.3) | 30(4.2) | 43(4.4) | 19(2.5) | 2(0.8) | 0(0.2) | 0(0) | 1 | 73(8.6) | 21(3.5) |
| EffTox | 1(3.7) | 6(5.3) | 19(8.7) | 26(9.8) | 26(10.4) | 15(8.2) | 4(4.5) | 1(2.1) | 1(1.3) | 0 | 52(20.2) | 21(16.1) |
| *Scenario 3* | | | | | | | | | | | | |
| Probit-joint | 0(3) | 0(3) | 6(3.2) | 27(4) | 44(4.4) | 20(2.6) | 2(0.8) | 0(0.2) | 0(0) | 1 | 71(8.4) | 22(3.6) |
| EffTox | 1(3.7) | 6(5.2) | 24(9.5) | 23(9.9) | 24(9.6) | 15(7.7) | 5(4.6) | 1(2.3) | 1(1.4) | 0 | 47(19.5) | 22(16) |
| *Scenario 4* | | | | | | | | | | | | |
| Probit-joint | 0(3) | 0(3) | 6(3.3) | 28(4.1) | 47(4.5) | 16(2.4) | 1(0.8) | 0(0.2) | 0(0) | 1 | 75(8.6) | 17(3.4) |
| EffTox | 6(6.0) | 21(9.0) | 40(14.1) | 27(11.8) | 4(6.0) | 1(3.6) | 0(2.0) | 0(0.8) | 0(0.5) | 0 | 31(17.8) | 1(6.9) |
| *Scenario 5* | | | | | | | | | | | | |
| BLRM | 0(3) | 1(3.1) | 3(3.3) | 15(3.7) | 32(4.1) | 26(4.1) | 17(3.6) | 5(2.7) | 1(1.3) | 0 | 58(7.7) | 23(5.2) |
| Probit-tox only | 0(3) | 0(3) | 1(3.1) | 10(3.5) | 33(4.2) | 37(4) | 14(2.2) | 2(0.8) | 1(0.3) | 0 | 67(8.2) | 17(3.3) |
| Probit-joint | 0(3) | 0(3) | 3(3.1) | 24(3.6) | 45(4.5) | 24(3.6) | 3(1.9) | 0(0.7) | 0(0.2) | 2 | 69(8.1) | 3(2.8) |
| EffTox | 4(5.6) | 12(6.2) | 31(10.7) | 37(12.8) | 12(8.3) | 3(4.9) | 1(3.1) | 0(1.5) | 0(1.0) | 0 | 15(13.2) | 1(5.6) |
| *Scenario 6* | | | | | | | | | | | | |
| BLRM | 0(3) | 2(3.2) | 5(3.4) | 24(4) | 41(4.3) | 21(3) | 5(3) | 2(1.7) | 0(0.6) | 0 | 41(4.3) | 28(7.7) |
| Probit-tox only | 0(3) | 0(3) | 2(3.2) | 21(3.9) | 45(4.5) | 26(3.2) | 5(1.3) | 1(0.3) | 0(0) | 0 | 45(4.5) | 32(4.8) |
| Probit-joint | 0(3) | 0(3) | 3(3.2) | 24(3.9) | 47(4.5) | 24(2.8) | 3(1) | 0(0.2) | 0(0) | 2 | 47(4.5) | 27(4) |
| EffTox | 3(4.9) | 12(6.5) | 35(12.3) | 33(12.2) | 12(8.0) | 3(4.9) | 1(2.8) | 0(1.4) | 1(0.7) | 1 | 12(8.0) | 6(9.8) |

[1] None: none of the dose levels is selected

For Scenario 1- 4, since they share the same dose-toxicity curve, the additional gain in terms of percentage of target dose levels being selected as MTD from incorporating the efficacy/immune biomarker can be evaluated. In general, the proposed method can achieve higher percentage of correctly select the target dose levels and maintaining slightly better overdosing control when efficacy/immune biomarker being introduced. The total mean number of patients being treated at over-toxic doses are lower which indicates that on average the Probit model with latent variable using overdose control level at 0.4 is generally safer as compared with BLRM using the same overdose control level. This amount of tolerance of overdosing of the proposed method is almost similar as BLRM using the overdose control level of 0.25. The proposed method shows obvious advantage over the toxicity alone method when the efficacy/immune biomarker shows difference between the dose candidates. The candidates with higher efficacy/immune biomarker responses would be more preferred in Scenario 3 and 5. Dose level 4 should be selected more often than dose level 5 in scenario 4, but since the dose-response curve estimation will have some delay, the estimated curve peak would occur later such that the estimated efficacy/immune biomarker response for dose level 5 would be higher than that of dose level 4. In scenario 6, the advantage of selecting the target dose is also increased. In general, the EffTox method has the lowest percentage of selecting the over-toxic doses, however the correctly selected percentage of target doses is also the lowest due its conservativeness.

For joint modeling toxicity and efficacy/immune biomarker, we consider model with additional biomarker. One choice of additional biomarker could be safety biomarker. For this type of biomarker, usually the lower the measurement is, the less the safety concern/toxicity is. For example, Cytokine Release Syndrome related biomarker. Another choice of biomarker/ measurement could be chosen that the lower the measurement is, the higher the safety concern/toxicity is, e.g., some biological measurements need to maintain above certain threshold to be normal. The cut-offs for scenarios 7, 8, and 9 are 0.3, 0.25, and 0.5 respectively. Simulation scenarios are listed in Table 4. Simulation results are summarized in terms of percentage of selection as MTD in Table 5.

**Table 4:** Simulation scenarios for three outcomes

| Dose | 60 | 75 | 90 | 105 | 120 | 135 | 150 | 165 | 180 | Target Dose Level |
|---|---|---|---|---|---|---|---|---|---|---|
| Dose Level | 1 | 2 | 3 | 4 | 5 | 6 | 7 | 8 | 9 | |

|  | 1 | 2 | 3 | 4 | 5 | 6 | 7 | 8 | 9 |  |
|---|---|---|---|---|---|---|---|---|---|---|
| Scenario 7 | | | | | | | | | | |
| Prob DLT | 0.01 | 0.05 | 0.10 | **0.18** | **0.27** | 0.38 | 0.5 | 0.55 | 0.7 | 4 |
| Prob Eff | 0.05 | 0.10 | 0.18 | **0.22** | **0.23** | **0.24** | **0.25** | **0.26** | **0.27** | |
| Prob Bio | **0.08** | **0.12** | **0.23** | **0.28** | 0.35 | 0.41 | 0.46 | 0.51 | 0.54 | |
| Scenario 8 | | | | | | | | | | |
| Prob DLT | 0.01 | 0.05 | 0.10 | **0.18** | **0.27** | 0.38 | 0.5 | 0.55 | 0.7 | 4, 5 |
| Prob Eff | 0.05 | 0.10 | 0.18 | **0.38** | **0.4** | **0.42** | **0.44** | **0.45** | **0.46** | |
| Prob Bio | 0.08 | 0.12 | 0.23 | **0.28** | **0.35** | **0.41** | **0.46** | **0.51** | **0.54** | |
| Scenario 9 | | | | | | | | | | |
| Prob DLT | 0.01 | 0.05 | 0.10 | **0.18** | **0.27** | 0.38 | 0.5 | 0.55 | 0.7 | 4 |
| Prob Eff | 0.05 | 0.10 | 0.18 | **0.28** | **0.4** | **0.42** | **0.44** | **0.45** | **0.46** | |
| Prob Bio | **0.80** | **0.78** | **0.65** | **0.55** | 0.43 | 0.35 | 0.30 | 0.2 | 0.2 | |

**Table 5:** Simulation results for three outcomes

| Dose Level | 1 | 2 | 3 | 4 | 5 | 6 | 7 | 8 | 9 | None[1] | Target Doses | Over-toxic doses |
|---|---|---|---|---|---|---|---|---|---|---|---|---|
| Scenario 7 | 1 | 2 | 4 | 45 | 36 | 10 | 1 | 0 | 0 | 1 | 45 | 11 |
| Scenario 8 | 0 | 1 | 7 | 36 | 40 | 13 | 2 | 0 | 0 | 1 | 76 | 15 |
| Scenario 9 | 0 | 0 | 7 | 45 | 34 | 12 | 1 | 0 | 0 | 1 | 45 | 13 |

[1]None: none of the dose levels is selected

The additional biomarker would provide additional gain in the percentage of selecting the targeting doses and reducing the risk of overdosing. However, the trend would be different when different cut-offs and corresponding dose relations present. Different weights can also be introduced to the model to balance the trade-off between different outcomes. For example, (Li et al. 2017) proposed to use joint unit probability mass with weight assigned to posterior probability of safety and efficacy endpoint through utility function. One can select the dose that has optimal safety and efficacy trade-off through the highest estimated posterior expected joint utility.

## 4. Discussion

We proposed a simple and flexible model that can integrate toxicity endpoint, efficacy biomarker information with correct usage of over-toxicity control in early phase dose escalation studies. The proposed latent Probit regression model can be easily extended to incorporate additional biomarkers or adjusted to fit study-specific biomarker usage. Our simulation studies shows that the proposed model has desirable operating characteristics with relative high selection rate of the target dose(s) that has(have) optimal risk-benefit trade-off and low selection rate of doses that are of high risk of over toxicity.

The biomarker information is utilized as binary or ordinal outcome, thus clinical/biological knowledge of the cut-offs value of biomarker measurement is crucial. Unlike the joint model proposed by Li et al. (2017), we model the toxicity endpoint and biomarker information directly in the joint multivariate normal distribution, which is more favorable when the underlying relation between different outcomes is vague but preliminary knowledge of biomarker still can support a threshold to differentiate clinical meaningful boundary, e.g., the rate of biomarker is classified as "response" is greater than 0.2. Thus, choice of thresholds is also a key point. The correlation between toxicity endpoint and biomarker information is modeled through the covariance matrix of the joint multivariate normal distribution. The covariance matrix is assumed to follow a Wishart distribution such that by adjusting the degree of freedom of Wishart distribution, one can assign different prior distribution to the covariance matrix, e.g., if 4 degrees of freedom is used, Wishart prior will put most weight on the extreme values $\rho_{ij} = 1$ or $-1$; when 3 degrees of freedom is applied, Wishart prior will have uniform distribution on interval $[-1,1]$ for $\rho_{ij}$ with $i \neq j$. When the underlying relations between different outcomes in the model are unclear, one can simply implement Wishart prior with 3 degrees of freedom for covariance matrix of the joint multivariate normal distribution for computational convenience. If other information on dose selection is available, our method could also be extended to a Bayesian approach that borrow information from historical data to provide more accurate estimation on parameters such as (Quan et al. 2021) and (Xu and Quan 2022). Models of biomarker and toxicity endpoints can be different in some cases, a joint model (e.g., Xu et al. 2021.) method may be used to catch the connection between two endpoints in more general way.

Compared with method that relying on toxicity alone, e.g., BLRM, the marginal performance gain of higher selection rate of target dose levels and lower selection rate of over-toxic dose levels comes from the additional biomarker information. When the biomarker dose-response curve quickly becomes plateau, because of the small sample size per dose level, the trend of changing slope cannot be picked up quickly by the model. Such that around the change point of dose-response curve, there would be some latency in terms of some slightly overestimates of the response rate for the dose levels right after the change point. If the target dose levels are somehow right after the change point of dose-response curve, the decision will be made as if the biomarker dose-response curve around target dose levels is increasing instead of plateau. In scenarios with bell shaped dose-response curve, if peak of dose-response curve is before target doses, the estimated dose-response curve's peak would occur slightly later, and shape may be flatter. This is reasonable, because of the overdosing control, subjects will less likely be assigned to higher dose levels. The dose-response estimate for high dose levels would more likely be influenced by prior.

Utility function can be used as ad-hoc safety and biomarker trade-off after the posterior estimates of the dose-toxicity and dose-biomarker response curves. However, the utility functions should be chosen carefully with reasonable clinical and medical justification. Usually, the utility score scale or utility function curve is not directly interpretable, and final dose recommendation can be sensitive to the choice of utility functions or scores. Compared with utility based BOIN12 (Lin et al. 2020) and two stage utility based U-BOIN (Zhou, Lee, and Yuan 2019), our proposed method's dose recommendation relies on direct inference on the full joint posterior distribution for the probability of DLT and biomarker response that satisfy certain clinical meaningful requirement, e.g., either interval or threshold, at each dose. Model free method, Ji3+3 (Lin and Ji 2020), is easy to implement, but the quadratic form of biomarker response curve in our model can capture a broad range of dose-biomarker relation, which allows investigate to incorporate different type of biomarkers or other measurement, e.g., normal vital signs rate at administration.

## Supplementary Materials

Additional simulation results for model with toxicity and efficacy biomarker.

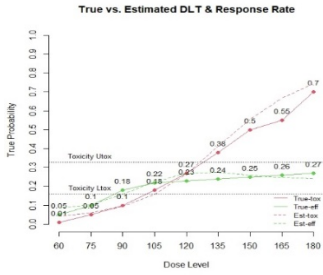
Scenario 1

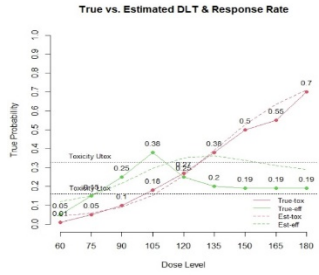
Scenario 2

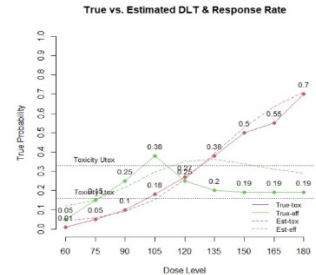
Scenario 3

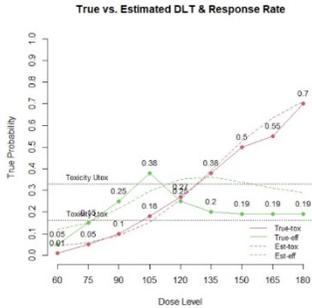
Scenario 4

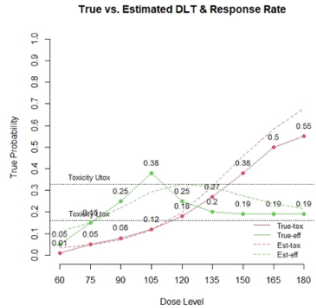
Scenario 5

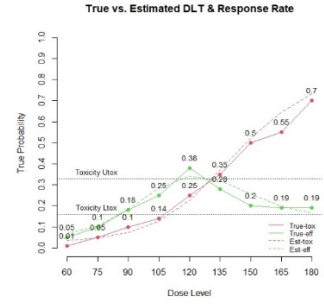
Scenario 6

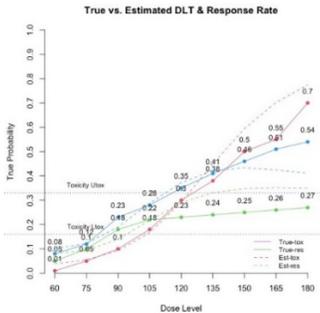
Scenario 7

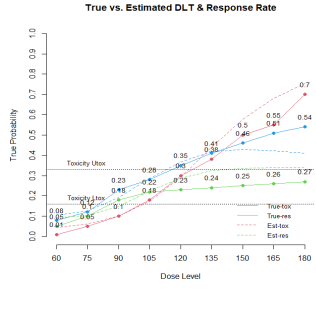
Scenario 8

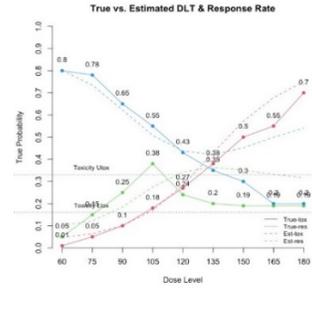
Scenario 9

**Figure S1:** Simulation result for all scenarios

Dashed lines in Figure S1 are estimated mean posterior dose-response/toxicity curves, solid lines are true curves.

**Table S1:** Simulation parameters for EffTox method

| Design Parameters | Description |
|---|---|
| *Acceptable dose* | |
| $\pi_T^* = 0.33 \ (= T_u)$ | Upper bound for acceptable probability of toxicity |
| $p_{T,L} = 0.6$ | Minimal acceptable risk for a dose to be safe (1- over dose control) |
| $\pi_E^* = 0.2 \ (= \theta_E)$ | Lower bound for acceptable probability of efficacy(activity) |
| $p_{E,L} = 0$ | Minimal risk of necessary efficacy(activity) for dose |
| *Trade-off* | |

| | | | | | | | | |
|---|---|---|---|---|---|---|---|---|
| $\pi^*_{1,E} = 0.50$ | | | Lower bound for acceptable probability of efficacy(activity) when 0% toxicity | | | | | |
| $\pi^*_{2,T} = 0.60$ | | | Upper bound for acceptable probability of toxicity when 100% activity | | | | | |
| $\pi^*_{3,E} = 0.70$ | | | Probability of efficacy(activity) and toxicity corresponding to an equivalent treatment as described above | | | | | |
| $\pi^*_{3,T} = 0.30$ | | | | | | | | |
| *Elicited means* | | | | | | | | |
| Dose level | 1 | 2 | 3 | 4 | 5 | 6 | 7 | 8 | 9 |
| Toxicity | 0.1 | 0.16 | 0.17 | 0.2 | 0.24 | 0.29 | 0.31 | 0.36 | 0.44 |
| Efficacy | 0.05 | 0.1 | 0.21 | 0.3 | 0.35 | 0.4 | 0.44 | 0.44 | 0.43 |

# References

Babb, J., A. Rogatko, and S. Zacks. 1998. "Cancer Phase I Clinical Trials: Efficient Dose Escalation with Overdose Control." *Statistics in Medicine* 17(10):1103–20.

Blumenthal, Gideon, Lokesh Jain, Anne Loeser, Yazadi K. Pithavala, Atiqur Rahman, Mark Ratain, Mirat Shah, Laurie Strawn, and Marc Theoret. 2022. "Optimizing Dosing in Oncology Drug Development Q & A." *FRIENDS of CANCER RESEARCH*. Retrieved October 3, 2022 (https://friendsofcancerresearch.org/wp-content/uploads/Optimizing_Dosing_in_Oncology_Drug_Development.pdf).

Colin, P., M. Delattre, P. Minini, and S. Micallef. 2017. "An Escalation for Bivariate Binary Endpoints Controlling the Risk of Overtoxicity (EBE-CRO): Managing Efficacy and Toxicity in Early Oncology Clinical Trials." *Journal of Biopharmaceutical Statistics* 27(6):1054–72.

Galon, Jérôme, Anne Costes, Fatima Sanchez-Cabo, Amos Kirilovsky, Bernhard Mlecnik, Christine Lagorce-Pagès, Marie Tosolini, Matthieu Camus, Anne Berger, Philippe Wind, Franck Zinzindohoué, Patrick Bruneval, Paul-Henri Cugnenc, Zlatko Trajanoski, Wolf-Herman Fridman, and Franck Pagès. 2006. "Type, Density, and Location of Immune Cells within Human Colorectal Tumors Predict Clinical Outcome." *Science (New York, N.Y.)* 313(5795):1960–64.

Greystoke, A., J. Cummings, T. Ward, K. Simpson, A. Renehan, F. Butt, D. Moore, J. Gietema, F. Blackhall, M. Ranson, A. Hughes, and C. Dive. 2008. "Optimisation of Circulating Biomarkers of Cell Death for Routine Clinical Use." *Annals of Oncology: Official Journal of the European Society for Medical Oncology* 19(5):990–95.

Guo, Beibei, Daniel Li, and Ying Yuan. 2018. "SPIRIT: A Seamless Phase I/II Randomized Design for Immunotherapy Trials." *Pharmaceutical Statistics* 17(5):527–40.

Li, Daniel H., James B. Whitmore, Wentian Guo, and Yuan Ji. 2017. "Toxicity and Efficacy Probability Interval Design for Phase I Adoptive Cell Therapy Dose-Finding Clinical Trials." *Clinical Cancer Research: An Official Journal of the American Association for Cancer Research* 23(1):13–20.

Lin, Ruitao, Yanhong Zhou, Fangrong Yan, Daniel Li, and Ying Yuan. 2020. "BOIN12: Bayesian Optimal Interval Phase I/II Trial Design for Utility-Based Dose Finding in Immunotherapy and Targeted Therapies." *JCO Precision Oncology* (4):1393–1402.

Lin, Xiaolei, and Yuan Ji. 2020. "The Joint I3+3 (Ji3+3) Design for Phase I/II Adoptive Cell Therapy Clinical Trials." *Journal of Biopharmaceutical Statistics* 30(6):993–1005.

Liu, Suyu, Beibei Guo, and Ying Yuan. 2018. "A Bayesian Phase I/II Trial Design for Immunotherapy." *Journal of the American Statistical Association* 113(523):1016–27.

Neuenschwander, Beat, Michael Branson, and Thomas Gsponer. 2008. "Critical Aspects of the Bayesian Approach to Phase I Cancer Trials." *Statistics in Medicine* 27(13):2420–39.

O'Quigley, J., M. Pepe, and L. Fisher. 1990. "Continual Reassessment Method: A Practical Design for Phase 1 Clinical Trials in Cancer." *Biometrics* 46(1):33–48.

Paoletti, Xavier, Christophe Le Tourneau, Jaap Verweij, Lillian L. Siu, Lesley Seymour, Sophie Postel-Vinay, Laurence Collette, Elisa Rizzo, Percy Ivy, David Olmos, Christophe Massard, Denis Lacombe, Stan B. Kaye, and Jean-Charles Soria. 2014. "Defining Dose-Limiting Toxicity for Phase 1 Trials of Molecularly Targeted Agents: Results of a DLT-TARGETT International Survey." *European Journal of Cancer (Oxford, England: 1990)* 50(12):2050–56.

Quan, Hui, Zhixing Xu, Meehyung Cho, Yingwen Dong, and Nan Jia. 2021. "Historical Control Data Borrowing for Noninferiority Assessment." *Statistics in Biopharmaceutical Research* 1–10.

Shah, Mirat, Atiqur Rahman, Marc R. Theoret, and Richard Pazdur. 2021. "The Drug-Dosing Conundrum in Oncology — When Less Is More." *New England Journal of Medicine* 385(16):1445–47.

Thall, Peter F., and John D. Cook. 2004. "Dose-Finding Based on Efficacy-Toxicity Trade-Offs." *Biometrics* 60(3):684–93.

Xu, Zhixing, and Hui Quan. 2022. "Bivariate Bayesian Hypothesis Testing with Missing Data in Components." *Pharmaceutical Statistics* 21(2):395–417.

Zhang, Wei, Daniel J. Sargent, and Sumithra Mandrekar. 2006. "An Adaptive Dose-Finding Design Incorporating Both Toxicity and Efficacy." *Statistics in Medicine* 25(14):2365–83.

Zhong, Wei, Joseph S. Koopmeiners, and Bradley P. Carlin. 2012. "A Trivariate Continual Reassessment Method for Phase I/II Trials of Toxicity, Efficacy, and Surrogate Efficacy." *Statistics in Medicine* 31(29):3885–95.

Zhou, Yanhong, J. Jack Lee, and Ying Yuan. 2019. "A Utility-Based Bayesian Optimal Interval (U-BOIN) Phase I/II Design to Identify the Optimal Biological Dose for Targeted and Immune Therapies." *Statistics in Medicine* 38(28):5299–5316.